# Objective Reality of de Broglie's Waves

Adriano Orefice*, Raffaele Giovanelli, Domenico Ditto

*Università degli Studi di Milano - DISAA - Via Celoria, 2 - 20133 - Milano (Italy)*

**Abstract -** An exact, ray-based general treatment is shown to hold for *any kind of monochromatic wave feature* - including diffraction and interference - described by Helmholtz-like equations, under the coupling action of a dispersive function (which we call "Wave Potential") encoded in the structure itself of the Helmholtz equation. Since the *time-independent* Schroedinger and Klein-Gordon equations (associating particles of assigned total energy with stationary *de Broglie waves*) are themselves Helmholtz-like equations, the same general approach is extended to the ray-based dynamics of point-like particles, whose *exact* trajectories and dynamical laws may be obtained *without resorting to statistical concepts of any kind*, thus suggesting a non-probabilistic nature of de Broglie's waves and of physical reality.

## 1 - Introduction

*"[La Mécanique Quantique], que je connais bien, puisque je l'ai longtemps enseignée, est très puissante et conduit à un très grand nombre de prévisions exactes, mais elle ne donne pas, à mon avis, une vue exacte et satisfaisante des phénomènes qu'elle étudie. Cela est un peu comparable au rôle joué naguère par la thermodynamique abstraite des principes qui permettait de prévoir exactement un gran nombre de phénomènes et était par suite d'une grande utilité, mais qui ne donnait pas une idée exacte de la réalité moleculaire dont le lois de la thermodynamique des principes ne donnaient que les conséquences statistiques".*
(Louis de Broglie,1972 [**1**])

The starting point of the present paper is the demonstration, performed in fully classical terms, that any kind of wave-like features may be treated (for monochromatic waves described by Helmholtz-like equations) by means of an exact, ray-based Hamiltonian kinematics, ruled by a dispersive function (which we call "*Wave Potential*") <u>encoded</u> in the structure of the Helmholtz equation, <u>avoiding</u> any statistical concept, <u>causing</u> a mutual coupling between monochromatic ray-trajectories and <u>representing</u> the one and only cause of any typically wave-like feature (such as diffraction and interference), while its omission leads to the usual geometrical optics approximation.

---

* Corresponding author  -  adriano.orefice@unimi.it

We extend, then, these properties to the case of Wave Mechanics, thanks to the fact that both the *time-independent* Schrödinger and Klein-Gordon equations (associating stationary *de Broglie waves* to particles of assigned total energy) are themselves Helmholtz-like equations, allowing to formulate the Hamiltonian dynamics of *point-like particles* in terms of *exact trajectories and motion laws*, under the coupling rule of a suitable "*Wave Potential*" in whose absence they reduce to the usual laws of classical dynamics.

While Bohm's "quantum trajectories" [**2**-**13**] may be viewed as probability flow-lines whose step-by-step construction requires the simultaneous solution of a *time-dependent* Schrödinger equation, our line of research [**14-16**], where no simultaneous solution of *time-dependent* equations is required, directly draws from the *time-independent* Schrödinger equation a set of *exact* point-particle trajectories of which Bohm's flow-lines represent a statistical average. The pre-eminent role of the *time-independent* Schrödinger equation is due to the fact that it's *the source, and not the consequence* (nor a minor particular case), of the *time-dependent* equation, inextricably accompanied by statistical concepts.

We give in Sects.2 and 3, respectively, the definition of the Wave Potential function in classical and quantum cases, leading to suitable, self-consistent sets of exact kinematic and/or dynamic equations and extend the theory, in Sect.4, to the relativistic case. We stress in Sects.5-6 the physical difference between the de Broglie-Bohm "causal" theories and the present one, and draw general conclusions in Sect.7, suggesting that Wave Mechanics describes - just like Classical Mechanics, of which it is a generalization - an objective reality.

**2- Hamiltonian ray-tracing of Helmholtz waves**

By assuming both wave monochromaticity and stationary isotropic media we sketch in the present Section an approach holding, in principle, for any kind of waves described by Helmholtz-like equations. In order to fix ideas we refer, within the present Section, to the case of classical electromagnetic waves of the form

$$\psi(\vec{r},\omega,t) = u(\vec{r},\omega)\, e^{-i\omega t}, \qquad (1)$$

where $\psi(\vec{r},\omega,t)$ represents any component of the electric and/or magnetic field and $u(\vec{r},\omega)$ is a solution of the Helmholtz equation [**17**]

$$\nabla^2 u + (n k_0)^2\, u = 0. \qquad (2)$$

The function $n(\vec{r},\omega)$ is the (time-independent) refractive index of the medium, and $k_0 \equiv \frac{2\pi}{\lambda_0} = \frac{\omega}{c}$. We now perform into eq.(2) the (quite general) well known replacement

$$u(\vec{r},\omega) = R(\vec{r},\omega)\, e^{i\varphi(\vec{r},\omega)}, \qquad (3)$$

with real $R(\vec{r},\omega)$ and $\varphi(\vec{r},\omega)$, which represent, respectively, *without any probabilistic meaning*, the amplitude and phase of the monochromatic wave.

Since the time dependence of the wave is known *a priori*, eq.(2) determines the exact *stationary* frame on which the wave-like features are distributed and where the ray propagation occurs. After the separation, in fact, of real and imaginary parts and the definition of the *wave vector*

$$\vec{k} = \vec{\nabla}\varphi(\vec{r},\omega) \tag{4}$$

and of the function

$$W(\vec{r},\omega) = -\frac{c}{2k_0}\frac{\nabla^2 R(\vec{r},\omega)}{R(\vec{r},\omega)}, \tag{5}$$

eq.(2) splits into the equation system

$$\begin{cases} \vec{\nabla}\cdot(R^2\vec{\nabla}\varphi) = 0 & (6) \\ D(\vec{r},\vec{k},\omega) \equiv \frac{c}{2k_0}[k^2 - (nk_0)^2] + W(\vec{r},\omega) = 0 & (7) \end{cases}$$

and the differentiation $\frac{\partial D}{\partial \vec{r}}\cdot d\vec{r} + \frac{\partial D}{\partial \vec{k}}\cdot d\vec{k} = 0$ of eq.(7) is seen to be satisfied by the Hamiltonian system

$$\begin{cases} \frac{d\vec{r}}{dt} = \frac{\partial D}{\partial \vec{k}} \equiv \frac{c\vec{k}}{k_0} & (8) \\ \frac{d\vec{k}}{dt} = -\frac{\partial D}{\partial \vec{r}} \equiv \vec{\nabla}[\frac{ck_0}{2}n^2(\vec{r},\omega) - W(\vec{r},\omega)] & (9) \end{cases}$$

associating with the Helmholtz equation (2) an *exact stationary set of trajectories* along which the rays - each one characterized by its launching position and wave vector - are driven, and providing also the ray motion laws $\vec{r}(\omega,t)$ and $\vec{k}(\omega,t)$ along the trajectories. A ray velocity $\vec{v}_{ray} = \frac{c\vec{k}}{k_0}$ is implicitly defined, and we may notice that, as long as $k \equiv |\vec{k}|$ remains equal to its launching value $k_0$, we'll have $v_{ray} \equiv |\vec{v}_{ray}| = c$. The function $W(\vec{r},\omega)$, which we call "*Helmholtz Wave Potential*", couples each trajectory to the adjacent ones in a kind of self-refraction, and is "encountered" by the rays, just like the refractive index $n(\vec{r},\omega)$, along their motion. Eq.(6), expressing the constancy of the flux of the vector field $R^2\vec{\nabla}\varphi$ along any tube formed by the field lines of the wave-vector $\vec{k} = \vec{\nabla}\varphi(\vec{r},\omega)$, plays a double role:

- <u>on the one hand,</u> since no new trajectory may suddenly arise in the space region spanned by the considered wave trajectories, and therefore $\vec{\nabla}\cdot\vec{\nabla}\varphi = 0$, eq.(6) tells us, when written in the explicit form $\vec{\nabla}\cdot(R^2\vec{\nabla}\varphi) \equiv 2R\vec{\nabla}R\cdot\vec{\nabla}\varphi + R^2\vec{\nabla}\cdot\vec{\nabla}\varphi = 0$, that $\vec{\nabla}R\cdot\vec{\nabla}\varphi = 0$: the amplitude $R(\vec{r},\omega)$ - together with its derivatives and functions, including $W(\vec{r},\omega)$ - is therefore distributed, at any step of the numerical integration, over the relevant wave-front, normal to $\vec{k} \equiv \vec{\nabla} \varphi(\vec{r},\omega)$, and the coupling term $\vec{\nabla} W(\vec{r},\omega)$ acts, at each point, *perpendicularly* to the ray trajectories. An interesting consequence of this property is the fact that, in the particular case of electromagnetic waves propagating *in vacuo* (i.e. for $n = 1$), the absolute value of the ray velocity $\vec{v}_{ray} = \dfrac{c\,\vec{k}}{k_0}$ remains equal to $c$ along each ray trajectory, whatever its form may be.

- <u>On the other hand</u>, eq.(6) provides step by step, after the assignment of the wave amplitude distribution $R(\vec{r},\omega)$ over the launching surface, the necessary and sufficient condition for the determination of $R(\vec{r},\omega)$ and $W(\vec{r},\omega)$ over the next wave-front, thus allowing a consistent "closure" of the Hamiltonian system. The choice of the launching distribution $R(\vec{r},\omega)$ has the role of describing the experimental set up, including for instance a suitable peak for every slit through which the wave enters the propagation region.

When, in particular, the space variation length $L$ of the wave amplitude $R(\vec{r},\omega)$ turns out to satisfy the condition $k_0 L \gg 1$, eq.(7) reduces to the well known *eikonal equation*

$$k^2 \equiv (\vec{\nabla}\varphi)^2 \simeq (n\,k_0)^2 \,. \qquad (10)$$

characterizing the so-called *geometrical optics approximation* [17]. In this approximation the rays are no longer mutually coupled by a Wave Potential, and propagate independently from one another under the only influence of the refractive index of the medium. The main consequence of this independence is *the absence, in such a limiting case, of typically wave-like phenomena such as diffraction and/or interference,* which may only be due to the coupling role of a non-vanishing Wave Potential.

It's worthwhile recalling that while the equation system (8-9) of the present paper provides an *exact* Hamiltonian description of the wave kinematics, an *approximate* ray-tracing (based on a *complex eikonal equation*, amounting to a first-order approximation of the wave beam diffraction) was presented in 1993/94 by one of the Authors (A.O., [18,19]), for the *quasi-optical* propagation of electromagnetic *Gaussian beams* at the electron-cyclotron frequency in the magnetized plasmas of Tokamaks such as JET and FTU, and applied in recent years [20] by an *équipe* working on the Doppler back-scattering microwave diagnostics installed on the Tokamak TORE SUPRA of Cadarache.

**3- Hamiltonian trajectories of de Broglie's waves**

Let us pass now to the case of non-interacting particles of mass $m$ and assigned total energy $E$ launched with an initial momentum $\vec{p}_0$ (with $p_0 = \sqrt{2\,m\,E}$) into a force field deriving from a stationary potential energy $V(\vec{r})$. The *classical* dynamical behavior of each particle may be described, as is well known [**17**], by the time-independent Hamilton-Jacobi equation

$$(\vec{\nabla} S)^2 = 2m[E - V(\vec{r})] \ , \qquad (11)$$

where the basic property of the function $S(\vec{r}, E)$ is that the particle momentum is given by

$$\vec{p} = \vec{\nabla} S(\vec{r}, E). \qquad (12)$$

In other words, the (time-independent) Hamilton-Jacobi surfaces $S(\vec{r}, E) = const$ are perpendicular to the momentum of the moving particles, and *pilot* them along *fixed trajectories* according to their dynamical motion laws.

One of the main forward steps in modern physics, giving rise to Wave Mechanics, was allowed by de Broglie's association of mono-energetic material particles [**21**, **22**] with suitable stationary "*matter waves*", according to the correspondence

$$\vec{p} / \hbar \equiv \frac{1}{\hbar} \vec{\nabla} S(\vec{r}, E) \rightarrow \vec{k} \equiv \vec{\nabla} \varphi, \qquad (13)$$

viewing the surfaces $S(\vec{r}, E) = const$, although *maintaining their piloting role and significance*, as the phase-fronts of the waves themselves. The relation (13) was suggested by considerations based on the Hamilton-Jacobi Dynamics and on the comparison between the variational principles of Maupertuis-Euler, $\delta \int_A^B p \, ds = 0$, holding in Dynamics, and Fermat, $\delta \int_A^B k \, ds = 0$, holding in Optics, together with the observation that the ratio between $p$ and $k$ has the dimensions of an *action* - just like the most famous microphysical quantity of our times, Planck's constant $\hbar$. The successive step was accomplished by Schrödinger [**23**, **24**], by assuming that Classical Mechanics (represented here by eq.(11)) be the *eikonal approximation of de Broglie matter waves*, and that these waves satisfy a Helmholtz equation of the form (2). By performing therefore into eq.(2) the replacement

$$(n k_0)^2 \cong k^2 \rightarrow p^2 / \hbar^2 \equiv (\vec{\nabla} \frac{S}{\hbar})^2 = \frac{2m}{\hbar^2}[E - V(\vec{r})] \ , \qquad (14)$$

allowed by eqs. (10)-(13), he got the *time-independent* equation

$$\nabla^2 u(\vec{r}, E) + \frac{2m}{\hbar^2}[E - V(\vec{r})] \, u(\vec{r}, E) = 0, \qquad (15)$$

holding [**25**, **26**] for the de Broglie waves associated with mono-energetic particles moving in a stationary potential field $V(\vec{r})$. Let us remind that the waves predicted by de Broglie were very soon confirmed by an experiment performed by Davisson and Germer on electron diffraction by a crystalline nickel target [**27**].

The same mathematical procedure applied in Sect.2 to the Helmholtz eq.(2) may now be applied to the Helmholtz-like Schrödinger equation (15), in order to search for a set of *exact particle trajectories*, corresponding to the ray trajectories of the previous Section. We put therefore in eq.(15), recalling eqs. (3) and (13),

$$u(\vec{r},E) = R(\vec{r},E)\, e^{\,i\,S(\vec{r},E)/\hbar}\ , \tag{16}$$

where the real functions $R(\vec{r},E)$ and $S(\vec{r},E)$ represent, respectively, the amplitude and phase of de Broglie's mono-energetic matter wave, whose objective reality is experimentally proven by diffraction and interference of particle beams. After separation of real and imaginary parts, and after having defined the function

$$Q(\vec{r},E) = -\frac{\hbar^2}{2m}\frac{\nabla^2 R(\vec{r},E)}{R(\vec{r},E)}, \tag{17}$$

we obtain the equation system

$$\begin{cases} \vec{\nabla}\cdot(R^2\,\vec{\nabla}S) = 0 & (18) \\ H(\vec{r},\vec{p},E) \equiv \dfrac{p^2}{2m} + V(\vec{r}) + Q(\vec{r},E) = E & (19) \end{cases}$$

to be compared with eqs.(6)-(7). The differentiation of eq.(19) leads then to the *dynamical* Hamiltonian system

$$\begin{cases} \dfrac{d\vec{r}}{dt} = \dfrac{\partial H}{\partial \vec{p}} \equiv \dfrac{\vec{p}}{m} & (20) \\ \dfrac{d\vec{p}}{dt} = -\dfrac{\partial H}{\partial \vec{r}} \equiv -\vec{\nabla}[V(\vec{r}) + Q(\vec{r},E)] & (21) \end{cases}$$

to be compared with the *kinematic* ray-tracing system (8)-(9). The *time-independent* Schrödinger equation (15) is therefore associated with a *stationary* set of trajectories along which point-like particles are *piloted* by de Broglie's mono-energetic waves according to their dynamical laws $\vec{r}(E,t)$ and $\vec{p}(E,t)$, *with no probabilistic implication*. The function $Q(\vec{r},E)$ of eq.(17) - which we call once more, for simplicity sake, *"Wave Potential"* - has the same basic structure and role of the Wave Potential function $W(\vec{r},\omega) = -\dfrac{c}{2k_0}\dfrac{\nabla^2 R(\vec{r},\omega)}{R(\vec{r},\omega)}$ of eq.(5): it has therefore *not so much a "quantum" as a "wave" origin*, entailed into quantum theory by de Broglie's waves. Just like the external potential $V(\vec{r})$, to which it's added, the *time-independent* Wave Potential $Q(\vec{r},E)$ is "encountered" by the particles along their motion $\vec{r}(E,t)$, and plays the basic role of mutually coupling the trajectories relevant to each mono-energetic matter wave. Once more, recalling the previous Section, the *presence* of the trajectory-coupling Wave Potential is the one and only cause of any wave-like feature**,** while its *absence* reduces the system (20)-(21) to the standard set of *classical dynamical* equations, which constitute therefore, as expected [**21-24**], its *geometrical optics* approximation. In complete correspondence with the electromagnetic case of Section 2,

- eq.(18) allows to obtain both $R(\vec{r},E)$ and $Q(\vec{r},E)$ along each trajectory, thus providing the "closure" of the quantum-dynamical system (20)-(21) and making a self-consistent numerical integration possible, without resorting to the simultaneous solution of any time-dependent Schrödinger equation, and

- the "force" term $\vec{\nabla}Q(\vec{r},E)$ due to the Wave Potential maintains itself *perpendicular* to the particle trajectories, so that *no energy exchange* is involved by its merely deflecting piloting action.

Many examples of numerical solution of the *dynamical* system (20)-(21) (or, equivalently, of the *kinematic* system (8)-(9)) in cases of diffraction and/or interference were given in Refs.[**14-16**], for simplicity sake, in the absence of external fields (or, in the case of eqs.(8)-(9), for *n =1* ) and in a geometry allowing to limit the computation to the *(x,z)*-plane, for waves launched along the *z*-axis. Referring, in order to fix ideas, to the *dynamical* system (20)-(21), the particle trajectories and the corresponding evolution both of the de Broglie wave intensity and Wave Potential were computed, with initial momentum components

$$p_x(t=0)=0; \quad p_z(t=0)=p_0=\hbar k_0 = 2\pi\hbar/\lambda_0, \qquad (22)$$

by means of a *symplectic* numerical integration method.

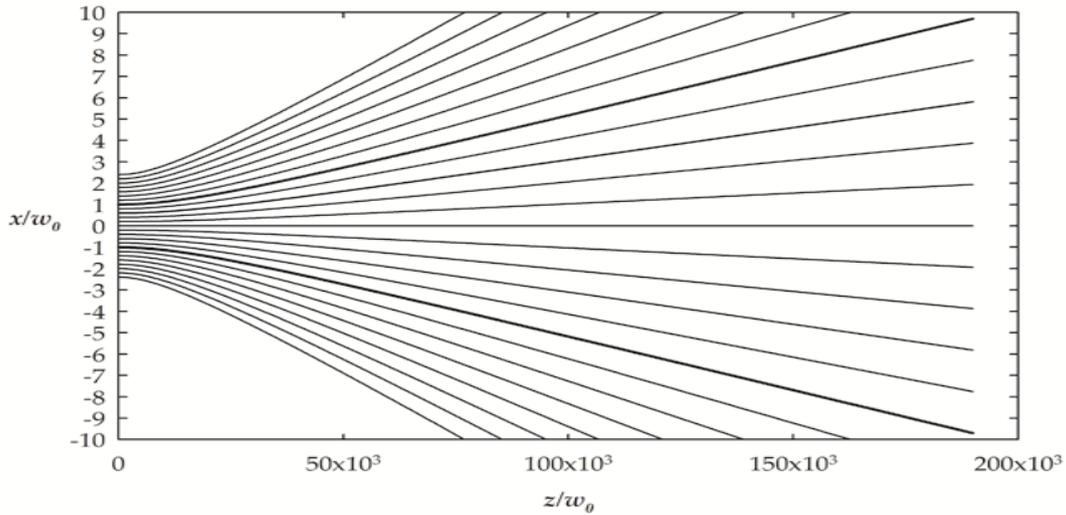

**Fig.1** Wave trajectories and waist lines on the symmetry *(x,z)*-plane for a *Gaussian* beam with waist $w_0$ and $\lambda_0/w_0 = 2\times 10^{-4}$ .

We limit ourselves to present here, in Fig.1, the particle trajectories on the *(x,z)*-plane relevant to the diffraction of a *Gaussian* particle beam traveling along *z* and starting from a vertical slit centered at $x=z=0$ in the form $R(x;z=0) \div exp(-x^2/w_0^2)$, where the length $w_0$ is the so-called *waist radius* of the beam. The two heavy lines are its *waist-lines*, given by the analytical relation

$$x(z)= \pm \sqrt{w_0^2 + \left(\frac{\lambda_0 \, z}{\pi \, w_0}\right)^2} \qquad (23)$$

representing, in the so-called *paraxial approximation* [**28**], the trajectories starting (at $z = 0$) from the *waist positions* $x = \pm w_0$ .

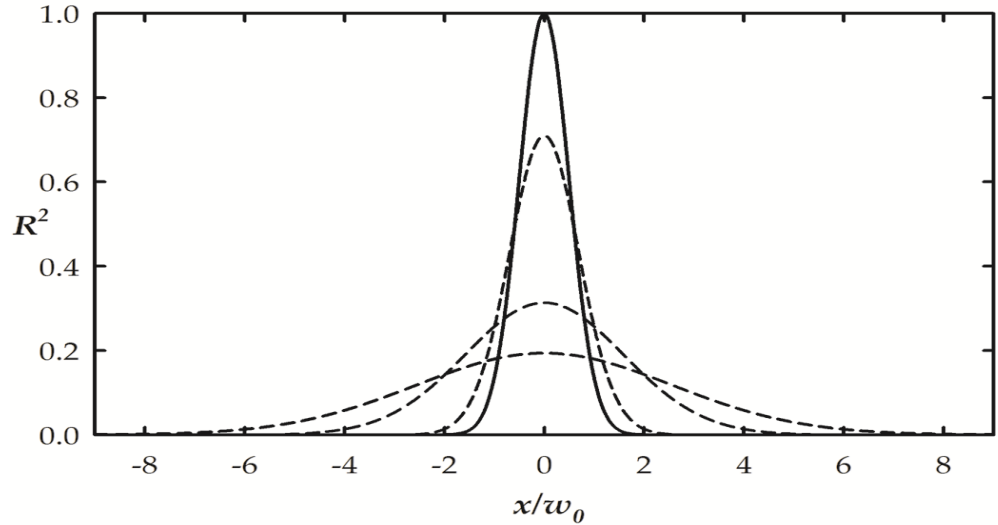

**Fig.2** Progressive flattening (dashed lines) at different times of the transverse intensity profiles of the de Broglie wave corresponding to Fig.1. The continuous line represent the initial profile.

The agreement between the *analytical* expression (23) and our *numerical* results provides, of course, an excellent test of our approach and interpretation. We present moreover, in Fig.2, the transverse intensity profiles, at different times, of the matter wave corresponding to Fig.1.

It's worthwhile reminding here that while our equations (20)-(21) provide an exact and general Hamiltonian description of the particle motion, an approximate treatment was presented in 1997 by one of the Authors (A.O., [**29**]), for the particular case of *Gaussian* particle beams. A *complex eikonal* equation, amounting to a first order approximation of the *quantum* particle diffraction, was adopted there - in complete analogy with the *classical* electromagnetic case of Refs.[**18**-**20**] - in order to overcome the collapse, for narrow beams, of the ordinary, zero-order, *real eikonal* approximation.

## 4 – Relativistic trajectories of de Broglie's waves

In order to extend the previous considerations to the relativistic case, we analyze now the motion of particles with rest mass $m_0$ and total energy $E$ traveling in a force field deriving from a static potential energy $V(\vec{r})$. Their behavior may be described by the relativistic *time-independent* Hamilton-Jacobi equation [**25**, **30**]

$$[\vec{\nabla} S(\vec{r}, E)]^2 = [\frac{E - V(\vec{r})}{c}]^2 - (m_0 c)^2 \tag{24}$$

which we interpret, once more, as the *eikonal approximation* of a de Broglie matter wave satisfying a Helmholtz-like equation of the form (2). By means of eqs. (10) and (24) we perform therefore de Broglie's replacement (13) into the Helmholtz eq.(2), in the form

$$(n k_0)^2 \cong k^2 \rightarrow p^2/\hbar^2 \equiv (\vec{\nabla}\frac{S}{\hbar})^2 \equiv [\frac{E-V(\vec{r})}{\hbar c}]^2 - (\frac{m_0 c}{\hbar})^2 \quad , \qquad (25)$$

reducing it to the *time-independent* Klein-Gordon equation

$$\nabla^2 u + [(\frac{E-V}{\hbar c})^2 - (\frac{m_0 c}{\hbar})^2] u = 0. \qquad (26)$$

The use, once more, of eq.(16), followed by the separation of real and imaginary parts, splits now eq.(26) into the equation system

$$\begin{cases} \vec{\nabla} \cdot (R^2 \vec{\nabla} S) = 0 \\ (\vec{\nabla} S)^2 - [\frac{E-V}{c}]^2 + (m_0 c)^2 = \hbar^2 \frac{\nabla^2 R(\vec{r},E)}{R(\vec{r},E)} \end{cases} \qquad (27)$$

Making use of the second of eqs.(27), and defining the function

$$H(\vec{r},\vec{p}) \equiv V(\vec{r}) + \sqrt{(pc)^2 + (m_0 c^2)^2 - \hbar^2 c^2 \frac{\nabla^2 R(\vec{r},E)}{R(\vec{r},E)}} = E \qquad (28)$$

we obtain the *dynamical* Hamiltonian system ("closed", as usual, by the first of eqs.(27))

$$\begin{cases} \frac{d\vec{r}}{dt} = \frac{\partial H}{\partial \vec{p}} \equiv \frac{c^2 \vec{p}}{E - V(\vec{r})} & (29) \\ \frac{d\vec{p}}{dt} = -\frac{\partial H}{\partial \vec{r}} \equiv -\vec{\nabla} V(\vec{r}) - \frac{1}{1 - V(\vec{r})/E} \vec{\nabla} Q(\vec{r},E) & (30) \end{cases}$$

where

$$Q(\vec{r},E) = -\frac{\hbar^2 c^2}{2 E} \frac{\nabla^2 R(\vec{r},E)}{R(\vec{r},E)} \quad . \qquad (31)$$

The system (29)-(30) provides the relativistic particle trajectories and motion laws, submitted once more to a mutual *Helmholtz coupling*, and reducing to the usual relativistic *dynamical* description in the absence of $Q(\vec{r},E)$ (i.e. in the *eikonal approximation* of de Broglie's waves). Once more, thanks to the first of eqs.(27), the term $\vec{\nabla} Q(r,E)$ acts *perpendicularly* to $\vec{p}$, whose amplitude cannot be modified by this coupling "force" of wave-like origin. Somewhat like in the case of a particle with electric charge *e* and relativistic mass *m* moving in *time-independent* electric and magnetic potentials $V(\vec{r})$ and $\vec{A}(\vec{r})$ (a case where $\vec{v} \neq \vec{p}/m$, because of the relation $\vec{p} = m\vec{v} + e\vec{A}(\vec{r})/c$), we have $\vec{v} \equiv \frac{d\vec{r}}{dt} \neq \vec{p}/m$ also in the present case, where however $\vec{v} \equiv \frac{d\vec{r}}{dt}$ is seen to maintain itself parallel to the momentum $\vec{p}$.

We conclude the present Section by observing that, in the particular case of *massless* particles (i.e. for $m_0 = 0$), the Klein-Gordon equation (26), by assuming the Planck relation

$$E = \hbar\omega, \tag{32}$$

takes on the classical-looking form

$$\nabla^2 u + (n\,\omega/c)^2\, u = 0, \tag{33}$$

with

$$n(\vec{r}, E) = 1 - V(\vec{r})/E. \tag{34}$$

Eq.(33) coincides with eq.(2), which may be therefore viewed as the time-independent Klein-Gordon equation holding for massless point-like particles in a stationary medium. We are brought back, then, to Sect.1 and to the ray trajectories found therein, with an insight of what is carried along by the "rays".

## 5 – Time-dependent vs time-independent Schrödinger equations

As we have shown in the previous Sections, the *time-independent* Schrödinger equation leads to the dynamical set of equations (20)-(21), providing the exact trajectories of point-like particles, *without resorting to any probabilistic conception*.

Let us now recall [25, 26] that, starting from eqs.(1) and (15), one may get the equation

$$\nabla^2 \psi - \frac{2m}{\hbar^2} V(\vec{r})\, \psi = -\frac{2m}{\hbar^2} E\, \psi \equiv -\frac{2mi}{\hbar}\, \frac{E}{\hbar\omega}\, \frac{\partial \psi}{\partial t}, \tag{35}$$

which, by assuming the Planck relation (32), reduces to the usual form of the *time-dependent* Schrödinger equation for a stationary potential field $V(\vec{r})$:

$$\nabla^2 \psi - \frac{2m}{\hbar^2} V(\vec{r})\, \psi = -\frac{2mi}{\hbar}\, \frac{\partial \psi}{\partial t}, \tag{36}$$

where $E$ and $\omega$ are not explicitly involved. Eq.(36), indeed, is not even a wave equation: its wave-like implications are due to the connection with the *time-independent* Schrödinger equation (15) from which it's obtained.

Eq.(15) admits in general, as is well known, a (discrete or continuous, according to the boundary conditions) set of energy eigen-values and orthonormal eigen-modes, which (referring for simplicity to the discrete case) we shall indicate, respectively, by $E_n$ and $u_n(\vec{r})$. It's a standard procedure to verify, making use of eqs.(1) and (32) and defining both the eigen-frequencies $\omega_n \equiv E_n/\hbar$ and the eigen-functions

$$\psi_n(\vec{r}, t) = u_n(\vec{r})\, e^{-i\omega_n t} \equiv u_n(\vec{r})\, e^{-iE_n t/\hbar}, \tag{37}$$

that a linear superposition (with constant coefficients $c_n$) of the form

$$\psi(\vec{r}, t) = \sum_n c_n\, \psi_n(\vec{r}, t), \tag{38}$$

is a general solution of eq. (36). Any component $\psi_n(\vec{r},t)$ corresponds to a single de Broglie wave, solving a *time-independent* Schrödinger equation (15) characterized by a different energy $E_n$, and the superposition (38) is subject to a spreading dispersion due to the divergence between the trajectories pertaining to different components. Although, at first sight, such a superposition simply represents, in duly normalized form, a weighted statistical average based on the available experimental information (represented by the coefficients $c_n$) about the relative weight of each $\psi_n(\vec{r},t)$, it owes its fame to Born's "ontological interpretation" [**31**] as the expression of *a physical state*, where energy is not determined: an interpretation which, even though "*no generally accepted derivation has been given to date*" [**32**], has become one of the standard principles of Quantum Mechanics and a new philosophical conception of an *intrinsically probabilistic* physical reality.

Let us also recall that while, starting from eq.(15), the *time-dependent* Schrödinger equation (36) is a mathematical truism, its "stronger" version,

$$\nabla^2 \psi - \frac{2m}{\hbar^2} V(\vec{r},t)\, \psi = -\frac{2mi}{\hbar}\, \frac{\partial \psi}{\partial t} \quad , \qquad (39)$$

containing a time-dependent external potential $V(\vec{r},t)$, may only be considered as an Ansatz, and is often assumed as a *First Principle*, enunciated at the very beginning of standard textbooks. Although it cannot be obtained from de Broglie's basic assumption (13), it was accepted by de Broglie himself [**33**] with these words: "*The form of* [eq.(36)] *allows us to go beyond single monochromatic waves and to consider superpositions of such waves. In addition, it suggests the way to extend the new Mechanics to the case of fields varying with time. Indeed, since it permits us to go beyond monochromatic waves, time no longer plays a special part, and it is then natural to* **admit** *that the form of the equation must be preserved when V depends on time as the general form of the equation of propagation of ψ waves in the non-relativistic Wave Mechanics of a single particle*". We limit ourselves to observe that eq.(39) cannot certainly lose the statistical character of eq.((36) from which it is *induced*.

**6 - Comparison with the de Broglie-Bohm "causal" theories**

Once established the different physical roles of the *time-independent* and *time-dependent* Schrödinger equations, let us pass now to consider Bohm's and de Broglie's "causal" approaches.

In **Bohm**'s approach a replacement of the form

$$\psi(\vec{r},t) = R(\vec{r},t)\, e^{\,i\, S(\vec{r},t)/\hbar} \qquad (40)$$

is performed into eq.(39) itself, splitting it, after separation of real and imaginary parts, into the equation system

$$\begin{cases} \dfrac{\partial P}{\partial t} + \vec{\nabla} \cdot (P \dfrac{\vec{\nabla} S}{m}) = 0 & (41) \\ \dfrac{\partial S}{\partial t} + \dfrac{(\vec{\nabla} S)^2}{2m} + V(\vec{r},t) - \dfrac{\hbar^2}{2m} \dfrac{\nabla^2 R}{R} = 0 & (42) \end{cases}$$

where, in agreement with the standard Copenhagen interpretation, the function $P(\vec{r},t) \equiv R^2(\vec{r},t)$ is assumed to represent the "*probability density for particles belonging to a statistical ensemble*", and eq.(41) is conceived as a *fluid-like* continuity equation for such a density probability.

Bohm's replacement (40) depicts the time-dependent function $\psi(\vec{r},t)$ as a single wave, *formally analogous* to de Broglie's (physically well-established) wave (16) and hopefully endowed with the same properties of experimental objective reality: it is, indeed, a strong attempt to dress with plausibility Born's interpretation of $\psi(\vec{r},t)$ as representing a "physical state". In order to develop this *formal analogy*, a basic Ansatz is performed by assuming that particles move according to the "*guidance formula*"

$$\vec{v}(\vec{r},t) = \vec{\nabla} S(\vec{r},t)/m, \qquad (43)$$

suggested by its *analogy* with the Hamilton-Jacobi momentum $\vec{p} \equiv \vec{\nabla} S(\vec{r}, E)$ *extended by de Broglie's seminal relation* (13) *to the exact dynamics of mono-energetic particles piloted by stationary matter waves*. A further *analogy* is found for eq.(42), which is viewed as a time-dependent Hamilton-Jacobi equation containing, in addition to the external potential $V(\vec{r},t)$, a "*Quantum* Potential"

$$Q_B(\vec{r},t) = -\dfrac{\hbar^2}{2m} \dfrac{\nabla^2 R(\vec{r},t)}{R(\vec{r},t)}, \qquad (44)$$

to be compared with the *Wave* Potential $Q(\vec{r},E) = -\dfrac{\hbar^2}{2m} \dfrac{\nabla^2 R(\vec{r},E)}{R(\vec{r},E)}$ of eq.(17). The time-integration, now, of the equation $\dfrac{d\vec{r}}{dt} = \vec{v}(\vec{r},t) = \vec{\nabla} S(\vec{r},t)/m$, to be performed in parallel with the simultaneous solution, for $R(\vec{r},t)$ and $S(\vec{r},t)$, of the *time-dependent* Schrödinger equation - associates with eq.(40) a set of so-called *quantum trajectories*, $\vec{r}(t)$, of the particles. Because, however, of the role itself of the *time-dependent* Schrödinger equation a probabilistic light is shed on $\vec{r}(t)$, so that Bohm's heuristic analogy appear to basically provide the flow-lines of the probability density $P(\vec{r},t)$.

Let us notice that, as observed by Bohm himself [**2**], one more *formal analogy* may be found in the classical Hamilton-Jacobi dynamical formulation holding in a *stationary state*, where the particle energy is given by the relation

$$E = -\dfrac{\partial S}{\partial t}, \qquad (45)$$

reducing the system (41)-(42) to the form

$$\begin{cases} \vec{\nabla} \cdot (R^2 \dfrac{\vec{\nabla} S}{m}) = 0 \\ -E + \dfrac{(\vec{\nabla} S)^2}{2m} + V(\vec{r}) - \dfrac{\hbar^2}{2m} \dfrac{\vec{\nabla} R}{R} = 0 \end{cases} \quad (46)$$

coincident with our equations (18)-(19), and containing a time-independent Potential of the form (17). While however our eqs.(20)-(21) are a direct consequence of de Broglie's basic relation (13), providing the exact, mono-energetic, *fine-grained* description of which any superposition (38) is a *coarse grained* average, Bohm's opportunity of discovering the *only possible case of exact particle trajectories* is missed by representing particles - even in the stationary case - by means of statistical ensembles.

We may also notice that the *time dependent* $Q_B(\vec{r},t)$ of eq.(44) presents in general, because of its structure, *non-local* properties due to the fact that it carries instant "in flight" information about any changes of the whole experimental arrangement: a behavior absent from the Wave Potential $Q(\vec{r},E)$ of eq.(17), which simply an equivalent expression of the stationary wave equation.

Coming now to the case of **de Broglie**'s "*double solution theory*" [**33-37**], we notice that the expression of the relativistic particle velocity given by eq.(29) of the present paper,

$$\vec{v} \equiv \dfrac{d\vec{r}}{dt} = \dfrac{c^2 \vec{p}}{E - V(\vec{r})}, \quad (47)$$

coincides with the *relativistic* "*guidance formula*" found by de Broglie, in time-independent fields, for the velocity of the (soliton-like) "*minute singular regions*" (representing physical particles) non-linearly included in the *objective* physical part of his *double solution*, in association with a *subjective* "*fictitious $\psi$ wave of statistical significance*". De Broglie bases his considerations, however, on a *time-dependent* Klein-Gordon equation, lending itself, in principle, to the same objections holding for Bohm's use of the time-dependent Schrödinger equation. Referring to a particle with electric charge $e$ moving in an electromagnetic field derivable from a scalar potential $U(\vec{r},t)$ and a vector potential $\vec{A}(\vec{r},t)$, de Broglie obtains the general "guidance formula"

$$\vec{v} = c^2 \dfrac{\vec{p} - e\vec{A}(\vec{r},t)/c}{-\partial S(\vec{r},t)/\partial t - eU(\vec{r},t)}, \quad (48)$$

describing the particle motion under the action of a "Quantum Potential" of the *statistical* form (44). Although eq.(48) *does* reduce, in stationary states (by making use of eq.(45) and assuming $\vec{A}=0$, $eU = V(\vec{r})$ ) to the form (47), the picture is hindered by the ambitious representation of the particle as a small clock, endowed with an internal oscillation in phase with the physical wave carrying it, and by the relation between this wave and the "subjective" statistical part of de Broglie's *double solution* - and did never reach the stage of trajectory calculation.

## 7 - Discussion and conclusions

As shown in Sect.2, an exact ray-based kinematics is encoded in the structure itself of Helmholtz-like equations, employed all over *Classical Mechanics* without any trace of probabilistic assumptions.

The observation, then, that both the time-independent Schrödinger and Klein-Gordon equations belong to the family of Helmholtz-like equations directly leads to attribute *exact* trajectories and motion laws to the particles associated with de Broglie's (objective and experimentally well established) stationary waves.

This is in conflict, of course, with the current idea that the very concept of exact particle trajectories is *always* physically meaningless: a conflict which is already present in Bohm's "causal" theory (when he says that "*precisely definable and continuously varying values of position and momentum*" [**2**] may be associated, in principle, with each particle) but is avoided, in practice, by castling on statistical ensembles of particles.

Although it is quite generally accepted [**38, 39**] that a *disturbance interpretation* of the uncertainty principle is nowadays untenable, one could distinguish, indeed, between the notions of *state preparation* of the experiment and of *state-disturbing* measurement [**40**]. Our "simple minded" objection is however: "why should an approach which is lawful and fruitful in Classical Mechanics be forbidden in the case of Wave Mechanics, if not for merely ideological reasons?".

We avoid therefore any distinction, and deem, in conclusion, that the present work provides the "missing link" between the "*exact*" description of particle motion characterizing the spirit itself of Classical Mechanics and the *probabilistic* Copenhagen description: a link presenting, <u>on the one hand</u>, de Broglie's stationary waves and their exact mono-energetic trajectories as an uncertainty-free extension of Classical Mechanics (allowed by the Helmholtz-like *time-independent* Schrödinger equation), and, <u>on the other hand</u>, Bohm's probability flow-lines and Schrödinger's *time-dependent* equation itself as the statistical treatment of this extended Mechanics.